\documentclass[]{scrartcl}
\usepackage{PackageManager}

\usepackage{amsmath}
\usepackage{amsfonts}
\usepackage{amssymb}

\usepackage{graphicx}

\usepackage{inputenc}
\usepackage[T1]{fontenc}

\KOMAoption{abstract}{true}
\KOMAoption{twocolumn}{off}
\KOMAoption{parskip}{half-}
\recalctypearea

\usepackage{hyperref}
\hypersetup{
	colorlinks  = true,
    allcolors = blue,
	linktocpage = true,
	breaklinks  = true,
}
\usepackage[
	comma,
	numbers,
	square,
    sort&compress,
	]{natbib}
\title{
MPC for Aquifer Thermal Energy Storage Systems Using ARX Models
}
\date{2025}

\author{
    Johannes van Randenborgh and Moritz Schulze Darup
    \thanks{
        J. van Randenborgh and M. Schulze Darup are with the Control and Cyberphysical Systems Group, Faculty of Mechanical Engineering, TU Dortmund University, Germany
        {\ttfamily\small \{<first name>.<last name>\}@tu-dortmund.de}.
        This manuscript is accepted for publication at the 16th IEEE/IAS International Conference on Industry Applications in São Sebastião, Brazil.
    }
}

\titlehead{
        © 2025 IEEE. Personal use of this material is permitted. Permission from IEEE must be obtained for all other uses, in any current or future media, including reprinting/republishing this material for advertising or promotional purposes, creating new collective works, for resale or redistribution to servers or lists, or reuse of any copyrighted component of this work in other works.
}
\begin{document}

\maketitle

\begin{abstract}
An aquifer thermal energy storage (ATES) can mitigate CO$_\textbf{2}$ emissions of heating, ventilation, and air conditioning (HVAC) systems for buildings.
In application, an ATES keeps large quantities of thermal energy in groundwater-saturated aquifers.
Normally, an ATES system comprises two (one for heat and one for cold) storages and supports the heating and cooling efforts of simultaneously present HVAC system components.
This way, the operation and emissions of installed and, usually, fossil fuel-based components are reduced.

The control of ATES systems is challenging, and various control schemes, including model predictive control (MPC), have been proposed.
In this context, we present a lightweight input-output-data-based autoregressive with exogenous input (ARX) model of the hybrid ATES system dynamics.
The ARX model allows the design of an output-based MPC scheme, resulting in an easy-to-solve quadratic program and avoiding challenging state estimations of ground temperatures.
A numerical study discusses the accuracy of the ARX predictor and controller performance.
\end{abstract}

\section{Introduction} \label{sec:introduction}
The Technology Executive Committee represents the policy arm of the Technology Mechanism of the United Nations Framework Convention on Climate Change, and it promotes technology practices and solutions related to sustainable buildings to serve the Paris Agreement~\cite{FrameworkConventiononClimateChange.2024}.
Despite its efforts, a thorough deployment of such practices for, e.g., building's heating, ventilation, and air conditioning (HVAC) systems remains.

Accordingly, underground thermal energy storage (UTES) provides a promising contribution to the agreement's fulfillment.
It can harness excess heat and cold from the environment for the air conditioning of the building.
With this, the UTES system lowers the operation of concomitant and fossil fuel-based auxiliary power units (APU), which are a part of the building's HVAC system~\cite{Lee.2013}.
Consequently, the total energy consumption of the HVAC system drops.

A subclass of UTES is aquifer thermal energy storage (ATES), which captivates with large storage capacities utilizing groundwater-saturated aquifers~\cite{Lee.2013}.
In application, ATES systems comprise (at least) two ATES, one for heat and one for cold.
For large emission mitigation effects, ATES systems significantly and dynamically provide energy to the energy demand of the building.
They beneficially pre-condition the temperature of the HVAC's process fluid before the APU or the building.
Thus, the energy consumption of the APU can be mitigated.

\subsection{Principles of ATES Operation} \label{sec:principles-of-ates-operation}
ATES systems store excess heat from summer in their storage for heating in winter.
In winter, cold is stored and used in summer to achieve comfortable room temperatures.
Storing thermal energy in the ATES results in local temperature changes in the ground.
A warm ATES keeps heated groundwater, and a cold ATES keeps cooled groundwater.

Pumps propagate the groundwater between the ATES through a heat exchanger (HX), which allows energy exchange between the building and the ATES.
The ATES systems' operational modes are heating, storing, and cooling, and they depend on the fluid flow between the two ATES.

An ATES system and building in heating mode are illustrated in Fig.~\ref{fig:ATES-principle}.
The complete piping system has two independent flow cycles, which interact via the HX and are located in the building and the ground.
A pump extracts groundwater from the warm ATES and delivers the water to the heat exchanger.
There, the groundwater exchanges energy with the process fluid of the building.
After this, the groundwater is injected into the cold ATES.
On the building side, the process fluid receives thermal energy from the groundwater and heats up.
Next, the second APU heats the fluid before entering the building.
Inside the building, the process fluid cools down and delivers heat to the rooms and occupants.
After leaving the building, the first APU further cools the fluid before entering the HX again.
For cooling, the flow direction between the storages flips, and the pumps are switched off during storing mode.
Further details about ATES's operation are discussed by Ref.~\cite{Lee.2013}.

\subsection{Control Approaches for ATES} \label{sec:control-approaches-for-ATES}
Literature presents control approaches for ATES systems that vary from the manual human interaction with the system~\cite{Duus.2021} to sophisticated predictive control, such as model predictive control (MPC)\cite{Rostampour.2017b, Rostampour.2016c}.
In general, the control objective is to maximize the contribution of ATES systems to the building's energy demand.
This section provides a brief literature review on MPC for ATES systems.

Reference~\cite{Rostampour.2016c} discloses a linear hybrid ATES system model including a heat pump as an APU.
The model is derived with first-principle equations and considers the on/off status of the heat pump.
A major assumption is perfect energy mixing, which leads to spatially constant ground temperatures.
A time-invariant constant considers heat loss in the subsurface.
The hybrid model yields a mixed-integer MPC algorithm.

Next, Ref.~\cite{Rostampour.2016} introduces a nonlinear hybrid ATES system model comprising various sub-components (i.e., boiler, chiller, heat pump, etc.).
The system is connected to several buildings.
An indirect data-driven approach increases model accuracy with parameter estimation for the thermal energy loss based on simulation results.
The nonlinear hybrid model results in a mixed-integer multi-dimensional polynomial nonlinear program, which is difficult to solve.
A follow-up publication by Ref.~\cite{Rostampour.2017b} lowers computational times by pre-defining operational modes depending on outside temperatures.

Lately, Ref.~\cite{vanRandenborgh.2024} presents a partial-differential-equation-based modeling approach for MPC.
The model focuses on the spatial distribution of temperature change in the ground.
To make up for the lack of temperature sensors in the ground, state estimation with an unscented Kalman filter and moving horizon estimator in Ref.~\cite{vanRandenborgh.2025} is proposed.
The authors implement a mixed-integer MPC algorithm.
Further, the designed controller focuses on the optimal and sustainable ATES system operation while minimizing operational cost.

\subsection{Contributions and Organization} \label{sec:contributions-and-organization}
A main challenge for an industry-wide deployment of MPC for ATES systems is to derive models in the sweet spot between model accuracy, capable of capturing most general dynamics, and the use for MPC, yielding fast-solving MPC algorithms.
Discussed literature, however, builds on slow-solving mixed-integer programs (see, e.g., \cite{Borelli.2017}).
Further, a required state estimator for the spatially distributed temperatures in the ground may not be successful in all geological circumstances.

This publication deals with the research question of whether input/output system identification of ATES systems can be beneficially used to avoid hybrid models and, thus, mixed-integer programs.
In addition, input/output systems can bypass state estimations of states that are difficult to measure and estimate (e.g., ground temperatures far from the borehole).
This publication discusses autoregressive with exogenous input (ARX) models, which are used to approximate the hybrid ATES system.
Beneficially, they yield fast-solving linear-quadratic MPC algorithms.

A brief introduction to ARX models, their identification, and predictors is given in Section~\ref{sec:preliminaries-on-arx-identification}.
Section~\ref{sec:ates-system-model} briefly presents the nonlinear hybrid ATES system model.
The accuracy of the derived ARX predictor is discussed in a numerical study in Section~\ref{sec:numerical-studies}.
At the end of this section, the controller performance with the ARX predictor is compared to the controller with the linearized model from Ref.~\cite{vanRandenborgh.2024}, including a moving horizon estimator from Ref.~\cite{vanRandenborgh.2025}.
The conclusion of this publication and a brief outlook are given in Section~\ref{sec:conclusion-and-outlook}.

\section{Preliminaries on ARX Identification} \label{sec:preliminaries-on-arx-identification}
The multi-input-multi-output (MIMO) ARX model,
\begin{equation} \label{eq:arx-model}
    \yb(t_{k}) = \Gbc(z) \ub(t_{k}) + \Hbc(z) \nub(t_{k}) \, ,
\end{equation}
is defined by the linear discrete-time transfer functions $\Gbc(z) \in \R^{p \times m}$ for inputs~$\ub(t_{k}) \in \R^{m}$ and $\Hbc(z) \in \R^{p \times p}$ for noise~$\nub(t_{k}) \in \R^{p}$, where $z$ refers to the backward shift operator (i.e., $z^{-1} \yb(t_{k}) = \yb(t_{k-1})$).
The discrete-time domain is indicated with $t_{k}$, which refers to time~$t_{k} = k \Delta t$, where $\Delta t$ is the discretization step size.
ARX model theory and identification may be extensively studied with, for instance, Ref. \cite{Ljung.1999, Isermann.1992.2} and \cite{Nelles.2020}.

An important ARX assumption is that the denominators of the input~$\Gbc(z)$ and noise transfer function~$\Hbc(z)$ are equal.
In other words, the denominator dynamics of the noise are equal to those of the process.
Further, the numerator of the noise transfer function~$\Hbc(z)$ is one.
When a system fulfills both assumptions,
\begin{align}
    \label{eq:ARX-assumptions}
    \Gbc(z) & = \frac{\Bbc(z)}{\Abc(z)} \quad \text{and} \quad \Hbc(z) = \frac{1}{\Abc(z)} \, ,
\end{align}
an ARX predictor may yield perfect predictions.
Note that $\Bbc(z)$ is the numerator of the input transfer function~$\Gbc(z)$, and $\Abc(z)$ is the denominator of the input~$\Gbc(z)$ and noise transfer function~$\Hbc(z)$.

The optimal MIMO ARX predictor describes the relationship between the expected system output~$\ybh(t_{k})$ and recorded system inputs~$\ub(t_{k-i})$ and outputs~$\yb(t_{k-i})$ with
\begin{equation} \label{eq:arx-mimo-model}
    \ybh(t_{k}) = \sum_{i=1}^{\sigma} \Bbc_{i} \ub(t_{k-i}) - \Abc_{i} \yb(t_{k-i}) \, .
\end{equation}
Here, $\Bbc_{i} \in \R^{p \times m}$ and $\Abc_{i} \in \R^{p \times p} \quad \forall i \in \{1, ..., \sigma \}$ refer to the ARX predictor parameters that are used for $\Bbc(z)$ and $\Abc(z)$.
The variable~$\sigma$ refers to the order of the ARX predictor and determines the number of past inputs and outputs.
Now, system identification determines the parameters~$\Abc_{i}$ and $\Bbc_{i}$ by, e.g., minimizing the Frobenius norm of the ARX prediction~$\Vbh$ and data~$\Vb$
\begin{equation} \label{eq:arx-least-squares}
   \Zb^{*} : = \arg\min_{\Zb} \| \Vbh - \Vb \|_{F}^{2} \text{ s.t. } \Vbh = \Mb \Zb \, ,
\end{equation}
where the matrices~$\Vb$ and $\Mb$ comprise measured or preprocessed data from $\Nc$ time steps, and
\begin{equation}
    \Zb = \left( \Bbc_{1}^{\top}, ..., \Bbc_{\sigma}^{\top}, - \Abc_{1}^{\top}, ..., - \Abc_{\sigma}^{\top} \right)^{\top}
\end{equation}
stores the ARX predictor parameters from Eq.~\eqref{eq:arx-mimo-model}.
The definition of $\Vb$ and $\Mb$ depends on the ARX identification method \cite{Ljung.1999, Nelles.2020, Isermann.1992.2} and will be presented in the following paragraphs.

A common ARX identification method is the least squares (LS) method, which determines the optimal ARX predictor parameters $\Zb^{*}$ by minimizing the prediction error with Eq.~\eqref{eq:arx-least-squares} \cite{Ljung.1999, Nelles.2020}.
Note that $\Vb$ and $\Mb$ now impose the MIMO ARX predictor from Eq.~\eqref{eq:arx-mimo-model}.

When the assumptions for ARX models in Eq.~\eqref{eq:ARX-assumptions} do not hold, the LS method may yield biased and non-consistent parameters~$\Zb^{*}$, mitigating the ARX predictor's accuracy.
According to Ref.~\cite{Nelles.2020}, a biased ARX predictor systematically deviates from the optimal (non-biased, consistent) ARX predictor.
However, when the predictor is consistent, the bias can be removed using an infinite number of data in the optimization problem~\eqref{eq:arx-least-squares} (or many data).
Though the LS method does not ensure a consistent predictor for stochastic noise signals, which makes the design of a non-biased and consistent ARX predictor difficult.

In the context of ATES systems, stochastic noise signals can be expected (e.g., from the temperature sensors), which disobey the ARX noise assumption~\eqref{eq:ARX-assumptions}.
Accordingly, the correlation functions least squares (COR-LS) method may be used instead of the LS method, as it guarantees consistent ARX predictor parameters for any stochastic noise signal \cite{Nelles.2020, Isermann.1992.2}.
This opens the opportunity to design a close-to-non-biased and consistent ARX predictor with a large amount of data.

The COR-LS method minimizes with Eq.~\eqref{eq:arx-least-squares} the prediction error of the correlation function,
\begin{equation}
    \label{eq:arx-corr-mimo-model}
    \thetab_{\ub \yb}(\kappa) = \sum_{i=1}^{\sigma} \thetab_{\ub \ub}(\kappa-i) \Bbc_{i}^{\top} - \thetab_{\ub \yb}(\kappa-i) \Abc_{i}^{\top} \, ,
\end{equation}
which is determined by multiplying $\ub(t_{k})$ on Eq.~\eqref{eq:arx-mimo-model} from the left side in a pre-processing step.
Note that $\thetab_{\ub \yb}(\kappa) \in \R^{m \times p}$ and $\thetab_{\ub \ub}(\kappa) \in \R^{m \times m}$ denote the cross-correlation between the system input~$\ub(t_{k})$ and output~$\yb(t_{k})$ and the auto-correlation of the system input~$\ub(t_{k})$.
Here, $\thetab_{\ub \yb}(\kappa)$ stands for
\begin{equation}
    \thetab_{\ub \yb}(\kappa) = \begin{pmatrix}
        \theta_{\ub_{1} \yb_{1}}(\kappa) & ... & \theta_{\ub_{1} \yb_{p}}(\kappa) \\
        \vdots & & \vdots \\
        \theta_{\ub_{m} \yb_{1}}(\kappa) & ... & \theta_{\ub_{m} \yb_{p}}(\kappa) \\
    \end{pmatrix} \, ,
\end{equation}
where $\theta_{\ub_{1} \yb_{1}}(\kappa)$ denotes the cross-correlation of the first element of the input~$\ub(t_{k})$ and the output~$\yb (t_{k})$.
$\thetab_{\ub \ub}(\kappa)$ is defined accordingly.

The correlation between two arbitrary discrete-time signals, $f(t_{k})$ and $g(t_{k})$, is defined by
\begin{equation} \label{eq:corr-approx}
    \theta_{fg}(\kappa) := \frac{1}{\Nc + 1} \sum_{k=1}^{\Nc-\|\kappa\|} f(t_{k}) g(t_{k+\kappa}) \, ,
\end{equation}
where $\kappa$ refers to a time shift between the signals \cite{Isermann.1992.2}.

For MIMO systems, the data matrix~$\Mb$ is given by
\begin{equation}
    \Mb^{\top} = \begin{pmatrix}
        \thetab_{\ub \ub}^{\top}(-P + \sigma - 1) & ... & \thetab_{\ub \ub}^{\top}(P - 1)  \\
        \vdots  & & \vdots \\
        \thetab_{\ub \ub}^{\top}(-P) & ... & \thetab_{\ub \ub}^{\top}(P - \sigma) \\
        \thetab_{\ub \yb}^{\top}(-P + \sigma - 1) & ... & \thetab_{\ub \yb}^{\top}(P - 1)\\
        \vdots  & & \vdots \\
        \thetab_{\ub \yb}^{\top}(-P)  & ... & \thetab_{\ub \yb}^{\top}(P - \sigma)
    \end{pmatrix}.
\end{equation}
$\Mb^{\top}$ has $2 \sigma (m+p)$ rows and $(2 P - \sigma + 1)m$ columns.
The parameter~$P$ can be arbitrary and sets an upper bound on the time shift~$\kappa$ considered in the pre-processing.
However, it must be verified that $\thetab_{\ub \yb}(\kappa)$ and $\thetab_{\ub \ub}(\kappa)$ are close to zero for $\| \kappa \| > P$ (see \cite{Isermann.1992.2}).
The data matrix~$\Vb \in \R^{(2 P - \sigma+1) m \times p}$ is given by
\begin{equation}
    \Vb = \begin{pmatrix}
        \thetab_{\ub \yb}^{\top}(-P + \sigma) & ... & \thetab_{\ub \yb}^{\top}(P)
    \end{pmatrix}^{\top} \, .
\end{equation}

Depending on the dataset size~$\Nc$ and chosen parameter~$P$, the COR-LS method introduces fewer equality constraints to the optimization problem~\eqref{eq:arx-least-squares} compared to the LS method.
This might speed up the solver runtime of the optimization problem.
In application, however, the pre-processing may take more time (depending on the dataset size~$\Nc$) than the solver runtime speedup.
However, the guarantee of a consistent ARX predictor is more compelling than speedups during the ARX identification process.

\section{ATES System Model} \label{sec:ates-system-model}
\begin{figure}[t]
    \centering
    \includegraphics[trim={0cm 11.55cm 0cm 0cm},clip]{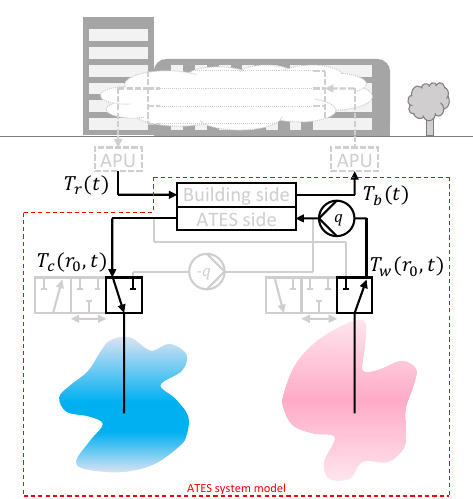}
    \caption{ATES system in heating mode with building and flow directions.
        The spatial locations of the system's output and input are illustrated.
        Components that are not comprised by the ATES system model are gray.
    }
    \label{fig:ATES-principle}
\end{figure}

Fig.~\ref{fig:ATES-principle} illustrates the considered ATES system in heating mode, which is connected to a building.
The Ref.~\cite{vanRandenborgh.2024} and \cite{vanRandenborgh.2025} fully disclose the mathematical model of this ATES system.
The red dotted line in Fig.~\ref{fig:ATES-principle} indicates the model boundaries.
Components, such as the building and the APU, are represented by gray dotted lines because they are not part of the ATES system model.
Further, subsystems, e.g., the valves, are given by gray solid lines when they are not active in heating mode but become active in other operational modes.
Valves switch the system between heating, storing, and cooling modes, and they indicate the hybrid nature of the system.

The nonlinear discrete-time piecewise state-space model,
\begin{align}
    \xb(t_{k+1}) & = \begin{cases}
        \Ab_{1}(q(t_{k})) \xb(t_{k}) + \Bb_{1} \ub(t_{k}) + \fb_{1} \text{ if } q(t_{k}) > 0 \\
        \Ab_{2}(q(t_{k})) \xb(t_{k}) + \Bb_{2} \ub(t_{k}) + \fb_{2} \text{ if } q(t_{k}) = 0 \\
        \Ab_{3}(q(t_{k})) \xb(t_{k}) + \Bb_{3} \ub(t_{k}) + \fb_{3} \text{ if } q(t_{k}) < 0 \\
    \end{cases} \nonumber \\
    \yb(t_{k}) & = \Cb \xb(t_{k}) \, ,
    \label{eq:ates-state-space-model}
\end{align}
represents the ATES system model.
The model's state~$\xb(t_{k})$ is composed of the HX outlet temperature at the building side~$T_{\mathrm{b}}(t_{k})$ and the spatially discretized ground temperature vectors~$\xb_{\mathrm{w}}(t_{k})$ and $\xb_{\mathrm{c}}(t_{k})$ for the warm and cold ATES (see \cite{vanRandenborgh.2024, vanRandenborgh.2025}).
The input~$\ub(t_{k}) \in \R^{m}$ comprises the groundwater flow~$q(t_{k})$ and the building return temperature~$T_{\mathrm{r}}(t_{k})$.
Further, the output matrix~$\Cb$ yields the output $\yb(t_{k}) \in \R^{p}$, which holds (in this order)
the HX outlet temperature at the building side~$T_{\mathrm{b}}(t_{k})$ and the temperature of the warm~$T_{\mathrm{w}}(r_{0},t_{k})$ and cold ATES~$T_{\mathrm{c}}(r_{0},t_{k})$ taken at the borehole with radius~$r_{0}$.
Please note that the return temperature~$T_{\mathrm{r}}(t_{k})$ is an exogenous input.
The state~$\Ab_{\{1, 2, 3\}}$ and input matrices~$\Bb_{\{1, 2, 3\}}$ and state offset vectors $\fb_{\{1, 2, 3\}}$ are defined for the operational modes: heating, storing, and cooling, i.e., $q(t_{k}) \, \{ >, =, <\} \, \qty{0}{\kilogram\per\second}$, respectively.

In fact, the ATES system model~\eqref{eq:ates-state-space-model} comprises three submodels, a model for the HX and the warm and cold ATES.
For the sake of completeness, a cursory overview of these submodels is given in the next paragraphs.

The HX is a central component connecting the ATES with the building and ensures the energy transfer between the ground and the building.
For heating mode, the outlet temperature at the ATES side of the cocurrent HX is given by this equation (see, e.g., \cite{Roetzel.2010})
\begin{multline}
    T_{\mathrm{c}}(r_{0}, t_{k + 1}) = \left(1 - \alpha_{\mathrm{a}}(t_{k}) \right) T_{\mathrm{w}}(r_{0}, t_{k}) \\
    + \alpha_{\mathrm{a}}(t_{k}) T_{\mathrm{r}}(t_{k}) \, ,
\end{multline}
where $\alpha_{\mathrm{a}}(t_{k})$ denotes the temperature change coefficient between the HX's inlet and outlet.
The outlet temperature at the building side~$T_{\mathrm{b}}(t_{k})$ is modeled accordingly.
For cooling mode, the ground temperatures of the warm and cold ATES change positions.

The ATES models consider temperature transport by a one-dimensional partial differential equation (PDE),
\begin{equation}
    \label{eq:T_PDE_nonlinear}
    c_{\mathrm{a}} \frac{{\partial T(r,t)}}{{\partial t}} =  \frac{\lambda}{r} \frac{\partial}{\partial r} \left( r \frac{\partial T(r,t)}{\partial r}\right) - \frac{ c_{\mathrm{w}} q(t_{k})}{2 \pi r l} \frac{\partial T(r,t)}{\partial r} \, ,
\end{equation}
which is slightly modified from the equation of energy \cite{Anderson.2005}.
Here, $c_{\mathrm{\{a,w\}}}$ denotes the specific volumetric heat capacity of the aquifer and groundwater, $\lambda$ the conduction coefficient of the ground, and $l$ the filter length of the ATES' boreholes.
The numerical solutions of PDE~\eqref{eq:T_PDE_nonlinear} yield the groundwater temperature over the radius~$r$ and time~$t$ (each in discrete steps).
In general, the PDE describes temperature changes based on heat conduction and convection effects that act simultaneously in the ground.
The specific boundary conditions to solve this PDE and further information are given by Ref.~\cite{vanRandenborgh.2024}.
Far from the borehole, a constant ambient temperature~$T_{\mathrm{amb}}$ is assumed.

The derived ATES system model generates data for the ARX identification presented in Section~\ref{sec:numerical-studies}.
The data is perturbed with Gaussian noise to emulate real-world process and measurement noise.

\section{Numerical Studies} \label{sec:numerical-studies}
The numerical studies are divided into three subsections.
First, the single-step prediction accuracy of the ARX predictor is examined in Section~\ref{sec:single-step-predictions-with-arx}.
Second, the prediction accuracy of the ARX predictor as a multi-step predictor is presented in Section~\ref{sec:multi-step-predictions-with-arx}.
Third, Section~\ref{sec:model-predictive-control-with-arx} discusses the performance of the ARX model with MPC.

The following paragraphs present the parameters of the model, which are valid for all subsections, except where the parameter is explicitly mentioned.
The study uses parameters related to an ATES system in Belgium.
Ref.~\cite{Vanhoudt.2011} and \cite{Desmedt.2007} provide a detailed description of the system.
Non-disclosed parameters are adequately chosen.

The cocurrent HX has a heat transfer coefficient of \qty[exponent-mode=input]{0.1}{\mega\watt\per\square\meter\kelvin} and a heat transfer area of \qty{10}{\square\meter}.
The process fluid has a specific volumetric heat capacity of $c_{\mathrm{w}} = \qty[exponent-mode=input]{4.2}{\mega\joule\per\cubic\meter\kelvin}$ and flows on both sides of the HX.
At the building side of the HX, the process fluid flow is constant with \qty[exponent-mode=input]{0.1}{\cubic\meter\per\second}.

The warm and cold ATES comprise rock and water with a volumetric heat capacity of $c_{\mathrm{a}} = \qty{4.4625}{\mega\joule\per\cubic\meter\kelvin}$, and the heat conduction coefficient~$\lambda$ ranges spatially with a uniform distribution between \num{3} and \qty{5}{\watt\per\meter\kelvin}.
The filter length is~$l = \qty{38}{m}$, and the borehole radius is $r_{0} = \qty[exponent-mode=input]{0.4}{m}$.
By assumption, the ambient temperature~$T_{\mathrm{amb}}$ remains at \qty{284.85}{\kelvin} (\qty{11.7}{\degreeCelsius}).
The upper and lower boundary for the pump flow~$q(t_{k})$ is set to $\pm$ \qty{100}{\cubic\meter\per\hour}.

The building return temperature~$T_{\mathrm{r}}(t_{k})$ is assumed given and ranges between \qty{276.15}{K} (\qty{3}{\degreeCelsius}) and \qty{291.15}{K} (\qty{18}{\degreeCelsius}).
In application, this temperature can be determined by measurements or by simulation.

All models are temporally discretized with $\Delta t = \qty{60}{s}$.
The warm and cold ATES model comprises a spatial domain with \num{15} cells, which ends at a distance of \qty{4}{\meter} from the boreholes.

The complete ATES system model consists of \num{33} states, two inputs, and three outputs (see Section~\ref{sec:ates-system-model}).
It is assumed that measurement and process noise follow a Gaussian distribution with zero mean and a standard deviation of \qty[exponent-mode=input]{0.3}{\kelvin}.

The training dataset holds $\Nc = \num[exponent-mode=input]{2900}$ simulatively generated samples.
Additionally, $\num{820}$ samples are simulated for a validation dataset.
The order of the ARX model is $\sigma = 3$, and the parameter $P$ for the ARX parameter identification with the COR-LS method is set to \num{50}.
Once identified, the ARX parameters~$\Abc_{i}$ and $\Bbc_{i}$ remain the same for all three subsections.

\subsection{Single-Step Predictions With the ARX Predictor} \label{sec:single-step-predictions-with-arx}
\begin{figure}[t]
    \centering
    \includegraphics[trim={0cm, 0.3cm, 0cm, 0cm}, clip, width=0.9\linewidth]{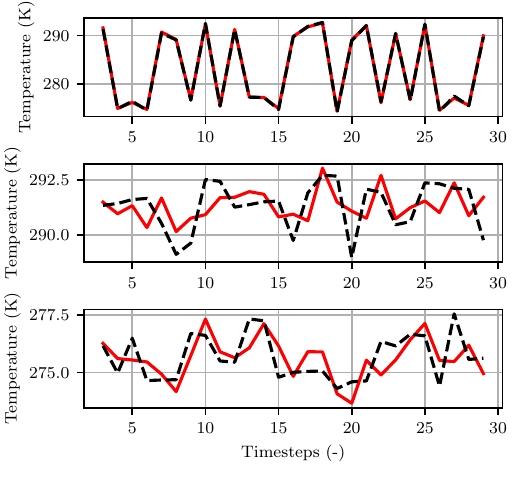}
    \caption{Predicted outputs of the ARX predictor with ARX parameters determined by the COR-LS (red) method in comparison to test data (black).
    }
    \label{fig:single-prediction}
\end{figure}

Figure~\ref{fig:single-prediction} compares the outputs of the ARX predictor with data from the validation dataset.
Note that Fig.~\ref{fig:single-prediction} shows a short sequence of $\num{27}$ data samples for a better illustration of the data.
The following statements about the prediction accuracy are based on \num{817} samples from the validation dataset.
The ARX predictor cannot generate predictions for the first samples because of its model order ($\sigma = 3$).

The mean prediction error, its standard deviation, and its maximal absolute error for all three outputs are disclosed in Tab.~\ref{tab:single-step}.
For all outputs, the mean prediction error is close to zero.
While the standard deviation of the HX outlet temperature~$T_{\mathrm{b}}(t_{k})$ is small, the standard deviation of the error for the other two outputs is about six times larger.
Obviously, the HX outlet temperature~$T_{\mathrm{b}}(t_{k})$ depends more on the last three system outputs and inputs than the ATES temperatures.
In application, the ground temperatures may vary in an interval of \qty{15}{\kelvin}.
Consequently, the maximal absolute error (MAE) for the ground temperatures climbs to about \qty{14}{\percent} of this interval.
This is reasonably small regarding the simplicity of the ARX model.

\begin{table}
    \caption{The Mean, Standard Deviation (STD), and Maximal Absolute Error (MAE) of Single-Step Predictions.
    }
    \centering
    \begin{tabular}{ c c c c }
         & \textbf{Mean} (\unit{\kelvin}) & \textbf{STD} (\unit{\kelvin}) & \textbf{MAE} (\unit{\kelvin}) \\
         $T_{\mathrm{b}}(\cdot)$ & \num{0.00557} & \num[exponent-mode=input]{0.13} & \num[exponent-mode=input]{0.47} \\
         $T_{\mathrm{w}}(\cdot)$ & \num{0.01037} & \num[exponent-mode=input]{0.81} & \num[exponent-mode=input]{2.69} \\
         $T_{\mathrm{c}}(\cdot)$ & \num{0.00662} & \num[exponent-mode=input]{0.80} & \num[exponent-mode=input]{2.46}
    \end{tabular}
    \label{tab:single-step}
\end{table}

\subsection{Multi-Step Predictions With the ARX Predictor} \label{sec:multi-step-predictions-with-arx}
This section discusses the multi-step accuracy of the ARX predictor in an MPC setting.
Section~\ref{sec:single-step-predictions-with-arx} shows that the ARX predictor provides single-step predictions with reasonable accuracy.
However, in the multi-step setting, the ARX predictor suffers from error accumulation.

For this numerical study, the prediction horizon is $N = 720$ time steps ($\widehat{=} \qty{12}{\hour}$), which aligns with the numerical study from Ref.~\cite{vanRandenborgh.2024}.
The validation dataset size allows for discussing the prediction accuracy with \num{100} samples, each holding \num{720} predictions for all three outputs.

Information about the prediction error for all samples is disclosed in Tab.~\ref{tab:multi-step}.
In comparison to Tab.~\ref{tab:single-step}, the mean error for the ground temperatures increased, which is due to error accumulations.
Further, the MAE error for the ground temperatures increased by $\approx \qty{1}{\kelvin}$ and represents, now, about \qty{25}{\percent} of the ATES temperature interval (\qty{15}{\kelvin}), which would suggest an intermediate prediction accuracy.

\begin{table}
    \caption{The Mean, the Mean Standard Deviation (STD), and Maximal Absolute Error (MAE) of Multi-Step Predictions.
    }
    \centering
    \begin{tabular}{ c c c c }
         & \textbf{Mean} (\unit{\kelvin}) & \textbf{STD} (\unit{\kelvin}) & \textbf{MAE} (\unit{\kelvin}) \\
         $T_{\mathrm{b}}(\cdot)$ & \num{0.00902} & \num[exponent-mode=input]{0.13} & \num[exponent-mode=input]{0.35} \\
         $T_{\mathrm{w}}(\cdot)$ & \num{0.74288} & \num[exponent-mode=input]{1.03} & \num[exponent-mode=input]{3.05} \\
         $T_{\mathrm{c}}(\cdot)$ & \num{0.94589} & \num[exponent-mode=input]{1.09} & \num[exponent-mode=input]{3.59}
    \end{tabular}
    \label{tab:multi-step}
\end{table}

\begin{figure}
    \centering
    \includegraphics[trim={0cm, 0.3cm, 0cm, 0cm}, clip, width=0.9\linewidth]{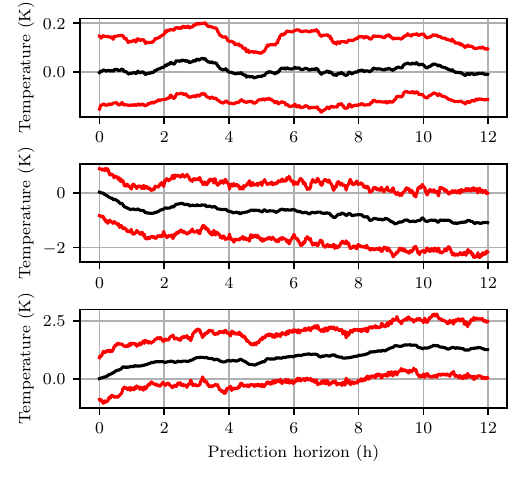}
    \caption{Mean prediction error (black) with its single standard deviation (red) interval over the prediction horizon~$N$.}
    \label{fig:multi-step}
\end{figure}

Nevertheless, plotting the mean error and its single standard deviation interval over the prediction horizon~$N$ in Fig.~\ref{fig:multi-step} reveals that the mean error of the ground temperatures drifts over the prediction horizon~$N$, and remains below \qty{1}{\kelvin} for the warm and cold ATES.
The standard deviation remains nearly constant over the prediction horizon~$N$.

\subsection{MPC With the ARX Predictor} \label{sec:model-predictive-control-with-arx}
The conditioning of buildings is a well-known application for MPC.
There, temperature constraints are usually interpreted as comfort limits, which may be (temporarily) exceeded.
Here, this is achieved by using soft constraints on the system outputs (see, for example, \cite{Ma.2012}).
Further, the input/output nature of the ARX predictor leads to an output-based MPC formulation.
The linear nature of the ARX predictor yields an easy-to-solve linear-quadratic MPC algorithm.

The optimal control problem (OCP) of the MPC scheme,
\begin{subequations} \label{eq:mpc-ocp}
    \begin{align}
        & \min_{\yb_{N}, \qb_{N}, \epsilonb_{N}} J(\yb_{N}, \qb_{N}, \epsilonb_{N}, t_{k}) \nonumber \\
        \text{subject to} \nonumber \\
        \underline{\qb}_{N} & \leq \qb_{N} \leq \overline{\qb}_{N} \, \label{eq:ocp-input-constraints}\\
        \underline{\yb}_{N} - \epsilonb_{N} & \leq \yb_{N} \leq \overline{\yb}_{N} + \epsilonb_{N} \label{eq:ocp-state-constraints}\\
        \zerob & \leq \epsilonb_{N} \label{eq:ocp-slack-constraints} \\
        \yb_{N} & = \Ab_{N} \yb_{N} + \Bb_{N} \qb_{N} + \fb_{N}(\yb_{\sigma}(t_{k})) \, , \label{eq:ocp-equality-constraints}
    \end{align}
\end{subequations}
minimizes the cost function~$J(\yb_{N}, \qb_{N}, \epsilonb_{N}, t_{k})$.
The OCP is constrained with the inequality constraints~\eqref{eq:ocp-input-constraints} on the pump flow~$\qb_{N} = (q(t_{k}), ..., q(t_{k+N-1}))^{\top}$ with upper and lower bounds~$\overline{\qb}_{N}, \underline{\qb}_{N}$, and soft constraints~\eqref{eq:ocp-state-constraints} on the system output~$\yb_{N} = (\yb^{\top}(t_{k}), ..., \yb^{\top}(t_{k+N}))^{\top}$ adding the slack variable~$\epsilonb_{N}$ (respecting Eq.~\eqref{eq:ocp-slack-constraints}) to the set of optimization variables.
Exceeding the upper and lower constraints, $\overline{\yb}_{N}$ and $\underline{\yb}_{N}$, leads to constraint violation costs.
The equality constraint~\eqref{eq:ocp-equality-constraints} predicts the future outputs~$\yb_{N}$ that are determined by the ARX predictor with the last recorded system outputs~$\yb^{\top}_{\sigma}(t_{k}) = \left( \yb^{\top}(t_{k}), ..., \yb^{\top}(t_{k-\sigma+1}) \right)$.
The optimizer of the OCP at time~$t_{k}$ leads to the optimal input~$q^{*}(t_{k})$ that is applied onto the system at time~$t_{k}$ until the new system output~$\yb(t_{k+1})$ is measured and re-initializes the OCP~\eqref{eq:mpc-ocp} with $\yb_{\sigma}(t_{k+1})$.

The quadratic cost function
\begin{equation*}
    J(\yb_{N}, \qb_{N}, \epsilonb_{N}, t_{k}) = \qb^{\top}_{N} \Rb_{N} \qb_{N} + \wb^{\top}_{N} \epsilonb_{N} + \left(\yb_{N} - \yb_{\mathrm{R}}(t_{k})\right)^{\top} \Qb_{N} \left(\yb_{N} - \yb_{\mathrm{R}}(t_{k})\right)
\end{equation*}
pushes the system towards predefined and time-dependent temperature setpoints~$\yb_{\mathrm{R}}(t_{k})$ for the complete prediction horizon~$N$.
At the same time, it minimizes the control input~$\qb_{N}$ and constraint violation costs~$\wb^{\top}_{N} \epsilonb_{N}$.
Each addend in the cost function is weighted with the cost weighting matrices, $\Rb_{N}$, $\wb_{N}$, and $\Qb_{N}$.

In this numerical study, the OCP of the MPC minimizes a multi-objective cost function to achieve the temperature setpoints~$\yb_{\mathrm{R}}(t_{k})$ and minimize the operational and constraint violation cost of the ATES system.
Here, temperature setpoints are only provided for the outlet temperature of the HX~$T_{\mathrm{b}}(t_{k})$, which are determined to fulfill the energy demand of the building.
The output bounds and cost weights are defined with
\begin{align*}
    \overline{\yb}_{N} & = \Ib_{N} \otimes \left(30, 25,11.7 \right)^{\top}, \, \underline{\yb}_{N} = \Ib_{N} \otimes \left(0, 11.7,0 \right)^{\top}, \\
    \Qb_{N} & = \Ib_{N}, \, \Rb_{N} = 0.01 \Ib_{N} \text{ and } \wb_{N} = \oneb_{p(N+1)} \, ,
\end{align*}
where $\otimes$ is the Kronecker product and $\Ib_{N}$ is the identity matrix with $N+1$ rows and columns.
$\oneb_{p(N+1)} \in \R^{p(N+1)}$ is the 1-vector.
$\overline{\yb}_{N}$ and $\underline{\yb}_{N}$ are given in degrees Celsius (\unit{\degreeCelsius}).

\begin{figure}
    \centering
    \includegraphics[trim={0cm, 0.3cm, 0cm, 0cm}, clip, width=0.9\linewidth]{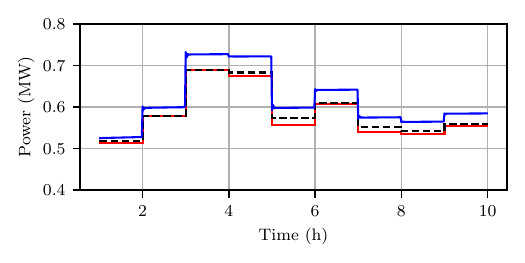}
    \caption{Delivered power by the ATES system with the MPC using the ARX predictor (blue) and the piecewise affine model from Ref.~\cite{vanRandenborgh.2024} and \cite{vanRandenborgh.2025} (red) compared to the energy demand (black, dashed) over time.}
    \label{fig:mpc}
\end{figure}

The closed-loop behavior of the MPC with the ARX predictor and the piecewise affine (PWA) model from Ref.~\cite{vanRandenborgh.2025, vanRandenborgh.2024} is compared.
Note that the MPC with the PWA model results in a mixed-integer quadratic program.
Figure~\ref{fig:mpc} illustrates the energy demand tracking results of the controller and compares them to the energy demand.
For comparison reasons, no process and measurement noise act on the feedback models.

MPC with the PWA model and a prediction horizon of $N = 720$ yields many integer optimization variables and, thus, an immense solver runtime ($>\qty{24}{\hour}$).
To avoid that, the discretization step size~$\Delta t$ was increased for the PWA MPC while lowering the prediction horizon~$N$.
With this, the prediction horizon remains equal in terms of time.
For PWA MPC, the prediction horizon is \num{12}, and the discretization step size is \qty[exponent-mode=input]{3600}{\second}.

The PWA MPC can follow the energy demand, as shown in red in Fig.~\ref{fig:mpc}.
In blue, the delivered energy demand by the MPC with the ARX predictor is illustrated.
The ARX MPC over-satisfies the energy demand of the building, which may be due to model mismatches between the ARX predictor and the nonlinear ATES system model.
In application, an over-delivery of the energy demand can usually be compensated for with other subsystems of the building or the environment.
For example, in winter, the warmer process fluid may be cooled by the cold environment.

Gurobi solves the quadratic program~\eqref{eq:mpc-ocp} with an average computational time of $\qty[exponent-mode=input]{0.07}{\second}$ \cite{GurobiOptimization.2023}.
This is an improvement compared to $>\qty{24}{\hour}$ and $\qty[exponent-mode=input]{2.3}{\second}$ (after adapting the discretization step size~$\Delta t$ and the prediction horizon~$N$) to solve the mixed-integer program of~\cite{vanRandenborgh.2024} on the same machine.

\section{Conclusion and Outlook} \label{sec:conclusion-and-outlook}
ATES systems may lower HVAC operation-related greenhouse gas emissions.
However, an optimal operation by sophisticated control is crucial for that to fully harness the emission mitigation potential.
One of the key challenges for an industry-wide integration of MPC is the model design, which should be capable of properly predicting system states and yielding fast-solving OCPs.
This paper discusses the potential of MIMO ARX predictors for the MPC of ATES systems to replace nonlinear and hybrid models for MPC.
Numerical studies examine the single- and multi-step ARX predictor accuracy and its performance with MPC.
The discussion of the MPC scheme in Section~\ref{sec:numerical-studies} shows that the ARX predictor lowers the solver runtime of the OCP drastically.

This ARX predictor and the autoregressive moving average with exogenous inputs building model of Ref.~\cite{Bunning.2022} may be combined for the MPC of a building with ATES system.


\begin{thebibliography}{10}

\bibitem{FrameworkConventiononClimateChange.2024}
{Technology Executive Committee}, ``{Rolling workplan of the Technology Executive Committee for 2023 - 2027}.'' https://unfccc.int/ttclear/tec/workplan, 2024.
\newblock Accessed: 02. May 2025.

\bibitem{Lee.2013}
K.~S. Lee, {\em {Underground thermal energy storage: Green energy and technology}}.
\newblock London, GBR: Springer, 2013.

\bibitem{Duus.2021}
K.~Duus and G.~Schmitz, ``{Experimental investigation of sustainable and energy efficient management of a geothermal field as a heat source and heat sink for a large office building},'' {\em Energy Build.}, vol.~235, 2021.

\bibitem{Rostampour.2017b}
V.~Rostampour, M.~Bloemendal, and T.~Keviczky, ``{A model predictive framework of ground source heat pump coupled with aquifer thermal energy storage system in heating and cooling equipment of a building},'' in {\em {Proc. 12th IEA Heat Pump Conf.}}, 2017.

\bibitem{Rostampour.2016c}
V.~Rostampour, M.~Jaxa-Rozen, M.~Bloemendal, and T.~Keviczky, ``{Building climate energy management in smart thermal grids via aquifer thermal energy storage systems},'' {\em Energy Procedia}, vol.~97, 2016.

\bibitem{Rostampour.2016}
V.~Rostampour, M.~Bloemendal, M.~Jaxa-Rozen, and T.~Keviczky, ``{A control-oriented model for combined building climate comfort and aquifer thermal energy storage system},'' in {\em {Proc. Eur. Geotherm. Congr.}}, 2016.

\bibitem{vanRandenborgh.2024}
J.~{van Randenborgh} and M.~{Schulze Darup}, ``{MPC using mixed-integer programming for aquifer thermal energy storages},'' {\em IFAC-PapersOnLine}, vol.~58, no.~18, 2024.

\bibitem{vanRandenborgh.2025}
J.~{van Randenborgh} and M.~{Schulze Darup}, ``{A moving horizon estimator for aquifer thermal energy storages}.'' https://arxiv.org/abs/2504.10203, 2025.
\newblock Accessed: 15. April 2025, in press.

\bibitem{Borelli.2017}
F.~Borelli, A.~Bemporad, and M.~Morari, {\em {Predictive control for linear and hybrid systems}}.
\newblock {Cambridge University Press}, 2017.

\bibitem{Ljung.1999}
L.~Ljung, {\em {System identification: Theory for the user}}.
\newblock {Prentice Hall information and system sciences series}, Upper Saddle River, NJ, USA: {Prentice Hall}, 2. ed.~ed., 1999.

\bibitem{Isermann.1992.2}
R.~Isermann, {\em {Identifikation dynamischer Systeme 2}}.
\newblock Berlin, DEU: Springer, 1992.

\bibitem{Nelles.2020}
O.~Nelles, {\em {Nonlinear System Identification}}.
\newblock Cham, CHE: Springer, 2020.

\bibitem{Roetzel.2010}
W.~Roetzel and B.~Spang, ``{Thermal design of heat exchangers},'' in {\em {VDI heat atlas}}, Berlin, DEU: Springer, 2010.

\bibitem{Anderson.2005}
M.~P. Anderson, ``{Heat as a ground water tracer},'' {\em Ground Water}, vol.~43, no.~6, 2005.

\bibitem{Vanhoudt.2011}
D.~Vanhoudt, J.~Desmedt, J.~{van Bael}, N.~Robeyn, and H.~Hoes, ``{An aquifer thermal storage system in a Belgian hospital: Long-term experimental evaluation of energy and cost savings},'' {\em Energy Build.}, vol.~43, no.~12, 2011.

\bibitem{Desmedt.2007}
J.~Desmedt, H.~Hoes, and N.~Robeyn, ``{Experiences on sustainable heating and cooling with an aquifer thermal energy storage system at a Belgian hospital},'' in {\em {Proc. Clima 2007 WellBeing Indoors}}, 2007.

\bibitem{Ma.2012}
Y.~Ma, A.~Kelman, A.~Daly, and F.~Borelli, ``{Predictive control for energy efficient buildings with thermal storage: Modeling, simulation, and experiments},'' {\em IEEE Control Syst.}, vol.~32, no.~1, pp.~44--64, 2012.

\bibitem{GurobiOptimization.2023}
{Gurobi Optimization L.L.C.}, ``{Gurobi optimizer reference manual}.'' https://www.gurobi.com, 2025.
\newblock Accessed: 10. Apr. 2025.

\bibitem{Bunning.2022}
F.~B{\"u}nning, B.~Huber, A.~Schalbetter, A.~Aboudonia, M.~{Hudoba de Badyn}, P.~Heer, and J.~Lygeros, ``{Physics-informed linear regression is competitive with two machine learning methods in residential building MPC},'' {\em Appl. Energy}, vol.~310, 2022.

\end{thebibliography}
\end{document}